\documentclass{article}
\usepackage[T1]{fontenc} % add special characters (e.g., umlaute)
\usepackage[utf8]{inputenc} % set utf-8 as default input encoding
\usepackage{ismir,amsmath,cite,url}
\usepackage{graphicx}
\usepackage{color}
\usepackage{enumitem}
\usepackage{float}
\usepackage{booktabs}
\usepackage{multirow}
\usepackage{amssymb}
\usepackage{tabularx}
\usepackage{booktabs}
\usepackage{hyperref}
\usepackage{caption}
\usepackage{subcaption}
\usepackage{circledsteps}
\usepackage{tikz}
\usepackage{titlesec}

\newcommand*\circled[1]{\tikz[baseline=(char.base)]{
    \node[shape=circle, draw, inner sep=1pt, 
        minimum height=12pt] (char) {#1};}}

\title{POP909: A POP-SONG DATASET FOR MUSIC ARRANGEMENT GENERATION}

\multauthor
{Ziyu Wang$^{1*}$ \hspace{1cm} Ke Chen$^{2*}$ \hspace{1cm} Junyan Jiang$^1$ \hspace{1cm} Yiyi Zhang$^3$} { \bfseries{\hspace{1cm} Maoran Xu$^4$ \hspace{1cm} Shuqi Dai$^5$ \hspace{1cm} Xianbin Gu$^1$ \hspace{1cm} Gus Xia$^1$}\\
 $^1$ Music X Lab, Computer Science Department, NYU Shanghai\\
  $^2$ CREL, Music Department, UC San Diego\\
  $^3$ Center for Data Science, New York University\\
  $^4$ Department of Statistics, University of Florida\\
    $^5$ Computer Science Department, Carnegie Mellon University\\
   $^*$ {\small The two authors have equal contribution.} \\
{\tt\small \{ziyu.wang, jj2731, yz2092, xianbin.gu, gxia\}@nyu.edu,} \\ 
{\tt\small knutchen@ucsd.edu, maoranxu@ufl.edu, shuqid@cs.cmu.edu}
}
\def\authorname{Ziyu Wang, Ke Chen, Junyan Jiang, Yiyi Zhang, Maoran Xu, Shuqi Dai, Xianbin Gu, Gus Xia}
\begin{document}

%s
\maketitle
\begin{abstract}
Music arrangement generation is a subtask of automatic music generation, which involves reconstructing and re-conceptualizing a piece with new compositional techniques. Such a generation process inevitably requires reference from the original melody, chord progression, or other structural information. Despite some promising models for arrangement, they lack more refined data to achieve better evaluations and more practical results. In this paper, we propose POP909, a dataset which contains multiple versions of the piano arrangements of 909 popular songs created by professional musicians. The main body of the dataset contains the vocal melody, the lead instrument melody, and the piano accompaniment for each song in MIDI format, which are aligned to the original audio files. Furthermore, we provide the annotations of tempo, beat, key, and chords, where the tempo curves are hand-labeled and others are done by MIR algorithms. Finally, we conduct several baseline experiments with this dataset using standard deep music generation algorithms.
\end{abstract}

\section{Introduction}\label{sec:intro}
\begin{figure}[t]
    \includegraphics[width = \columnwidth]{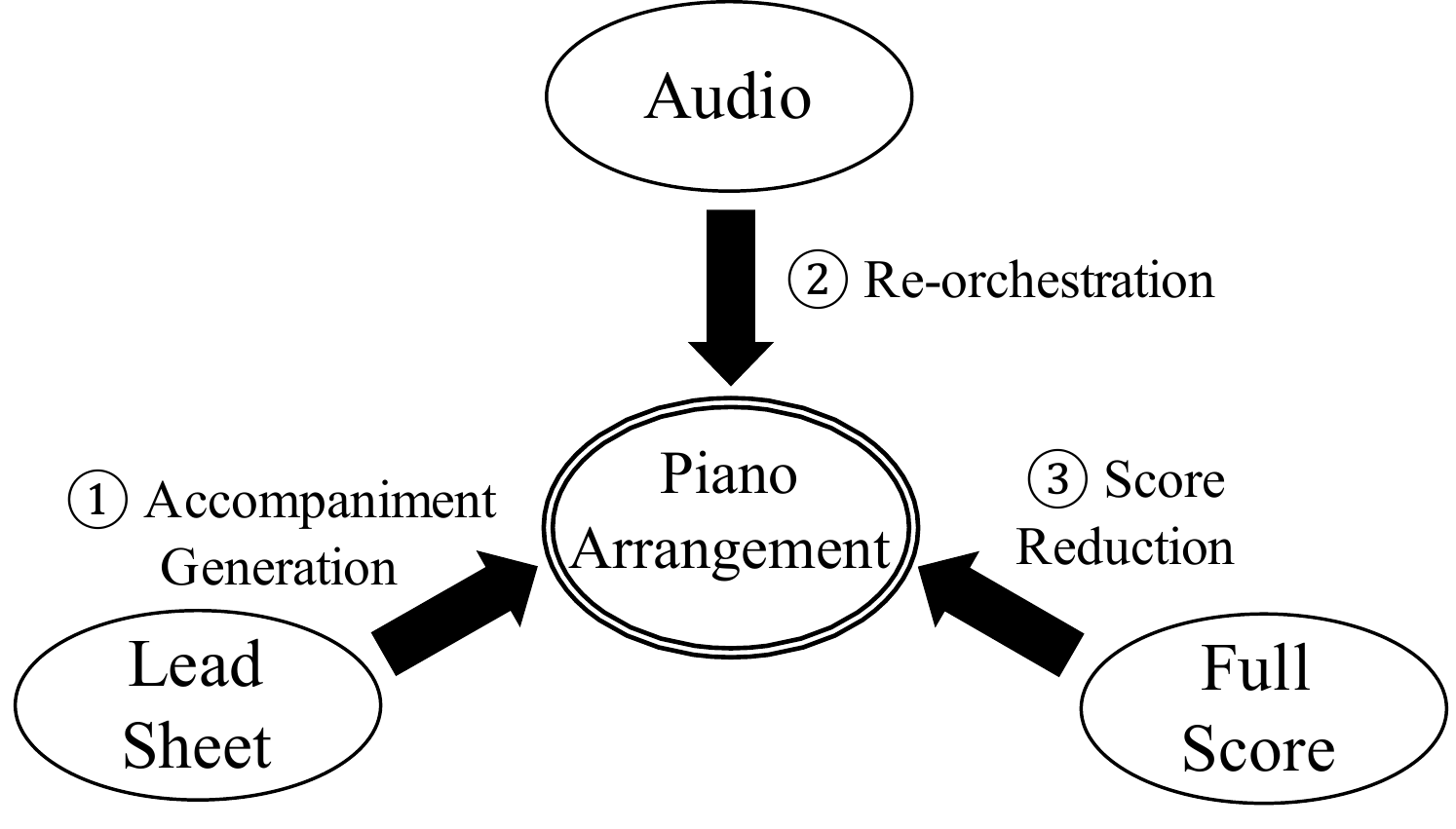}
    \caption{Illustration of the role of piano arrangement in the three forms of music composition, where \protect\circled{1} and \protect\circled{2} are covered by our POP909 dataset.
    }
    \label{fig:triangle}
\end{figure}

% para: 1. State the center position of piano arrangement among score, audio and full score.
Music arrangement, the process of reconstructing and re-conceptualizing a piece, can refer to various \textit{conditional} music generation tasks, which includes \textit{accompaniment generation} conditioned on a lead sheet (the lead melody with a chord progression)~\cite{mysong, elowsson, wang2018framework, musegan}, transcription and \textit{re-orchestration} conditioned on the original audio~\cite{audioreduction, audioreduction2, hung2019musical}, and \textit{reduction} of a full score so that the piece can be performed by a single (or fewer) instrument(s)~\cite{reduction1, reduction2}. As shown in Figure~\ref{fig:triangle}, arrangement acts as a bridge, which connects  lead sheet, audio and full score. In particular, \textit{piano arrangement} is one of the most favored form of music arrangement due to its rich musical expression. With the emergence of player pianos~\cite{mr} and expressive performance techniques \cite{xiaexpressive, jeong2019graph}, we expect the study of piano arrangement to be more meaningful in the future, towards the full automation of piano composition and performance.

\begin{table*}[]
\vspace{-0.4cm}
\centering
{\small
\begin{tabular}{|c|c|c|c|c|c|c|c|c|c|}
\hline
\multirow{2}{*}{\textbf{Dataset}} & \multirow{2}{*}{\textbf{Size}} & \multicolumn{4}{c|}{\textbf{Paired Property}} & \multicolumn{3}{c|}{\textbf{Annotation}} & \multirow{2}{*}{\textbf{Modality}}\\ \cline{3-9} 
&   & \textbf{Polyphony}   & \textbf{Lead Melody}  & \textbf{Audio} & \textbf{Time-alignment} & \textbf{Beat}    & \textbf{Key}   & \textbf{Chord} & \\ \hline
Lakh MIDI\cite{lakhmidi} & 170k & \checkmark   &   & \checkmark     & $\Delta$      & \checkmark       &  $\Delta$     & $\Delta$   & score, perf\\ \hline
JSB Chorales\cite{JSB-chorale}  & 350+    & \checkmark           &              &                 & N/A             & \checkmark       & \checkmark     &       & score        \\ \hline
Maestro\cite{maestro}                 & 1k                    & \checkmark           &              & \checkmark               &                 &         &       &            & perf   \\ \hline
CrestMuse\cite{crestmuse}                & 100                   & \checkmark           &              & \checkmark               & \checkmark               & \checkmark       & \checkmark     &             & score, perf  \\ \hline
RWC-POP\cite{rwcpop}                 & 100                   & \checkmark           & \checkmark            & \checkmark               & $\Delta$                & \checkmark       & \checkmark     & \checkmark         & score, perf    \\ \hline
Nottingham\cite{nottingham}               & 1k                    &             & N/A          &                 & N/A             & \checkmark       & \checkmark     & \checkmark         &    score\\ \hline
POP909                   & 1k                    & \checkmark           & \checkmark            & \checkmark               & \checkmark               & \checkmark       & \checkmark     & \checkmark  & score, perf           \\ \hline
\end{tabular}
}
\caption{A summary of existing datasets.}
\label{tab:dataset_sum}
\vspace{-0.4cm}
\end{table*}

In the computer music community, despite several promising generative models for arrangement, the lack of suitable datasets becomes one of the main bottlenecks of this research area (as pointed by\cite{musictransformer,lakhnes}.) A desired arrangement dataset should have three features. First, the arrangement should be a style-consistent re-orchestration, instead of an arbitrary selection of tracks from the original orchestration. Second, the arrangement should be paired with an original form of music (audio, lead sheet, or full score) with precise time alignment, which serves as a supervision for the learning algorithms. Third, the dataset should provide external labels (e.g., chords, downbeat labels), which are commonly used to improve the controllability of the generation process \cite{chen2019effect}. Until now, we have not seen such a qualified dataset. Although most existing high-quality datasets (e.g., \cite{lakhmidi,maestro}) contain at least one form of audio, lead melody or full score data, they have less focus on arrangement, lacking accurate alignment and labels. % In this study, we want fill this research gap in \circled{1} and \circled{2} arrangement tasks (indicated in Figure~\ref{fig:triangle}). 

To this end, we propose POP909 dataset.\!\footnote{The dataset is available at \href{https://github.com/music-x-lab/POP909-Dataset}{https://github.com/music-x-lab/POP909-Dataset}} It contains 909 popular songs, each with multiple versions of piano arrangements created by professional musicians. The arrangements are in MIDI format, aligned to the lead melody (also in MIDI format) and the original audios. Furthermore, each song are provided with manually labeled tempo curves and machine-extracted beat, key and chord labels using music information retrieval algorithms. We hope our dataset can help with future research in automated music arrangement, especially task \circled{1} and \circled{2} indicated in Figure~\ref{fig:triangle}:
\begin{itemize}[leftmargin=*]
    \item[] \textbf{Task 1: Piano accompaniment generation} conditioned on paired melody and auxiliary annotation. This task involves learning the intrinsic relations between melody and accompaniment, including the selection of accompaniment figure, the creation of counterparts and secondary melody, etc.
    \item[] \textbf{Task 2: Re-orchestration from audio}, i.e., the generation of piano accompaniment based on the audio of a full orchestra.
\end{itemize}
Besides those main tasks, our dataset can also be used for unconditional symbolic music generation, expressive performance rendering, etc.

\section{Related Work}
\label{sec:2:relatedwork}
In this section, we begin with a discussion of different modalities of music data in Section~2.1. We then review some existing composition-related datasets in Section 2.2 and summarize the requirements of a qualified arrangement dataset in Section~2.3. Again, our focus is piano arrangement and this dataset is designed for task \circled{1} and \circled{2} indicated in Figure~\ref{fig:triangle}, i.e., \textit{piano accompaniment generation} based on the lead melody or the original audio.

\subsection{Modalities of Music Generation}
As discussed in \cite{dai2018music}, music data is intrinsically multi-modal and most generative models focus on one modality. In specific, music generation can refer to: 1) \textit{score generation}~\cite{deepbach,musicinpaint,folkrnn,ICSC-Ke}, which deals with the very abstract symbolic representation, 2) \textit{performance rendering}~\cite{lakhnes,musictransformer,musegan}, which regards music as a sequence of controls and usually involves timing and dynamics nuances, and 3) \textit{audio synthesis}~\cite{wavenet,jukebox}, which considers music as a waveform or spectrogram. The POP909 dataset is targeted for arrangement generation in the modality of score and performance.

\subsection{Existing Datasets}
\begin{spacing}{1.02}
Table~\ref{tab:dataset_sum} summarizes the existing music datasets which are the potential resources for the piano arrangement generation tasks. The first column shows the dataset name, and the other columns show some important properties of each dataset.

Lakh MIDI \cite{lakhmidi} is one of the most popular datasets in symbolic format, containing 176,581 songs in MIDI format from a broad range of genres. Most songs have multiple tracks, most of which are aligned to the original audio. However, the dataset does not mark the lead melody track or the piano accompaniment track and therefore cannot be directly used for piano arrangement.

Maestro \cite{maestro} and E-piano \cite{e-piano} contains classical piano performances in time-aligned MIDI and audio formats. However, the boundary between the melody and accompaniment is usually ambiguous for classical compositions. Consequently, the dataset is not suitable for the arrangement task \circled{1}. Moreover, the MIDI files are transcription rather than re-harmonization of the audio, which makes it inappropriate for the arrangement task \circled{2} either.

Nottingham Database \cite{nottingham} is a high-quality resource of British and Irish folk songs. The database contains MIDI files and ABC notations. One drawback of the dataset is that it only contains monophonic melody without polyphonic texture.

RWC-POP \cite{rwcpop}, CrestMuse \cite{crestmuse}, and JSB-Chorale \cite{JSB-chorale} all contain polyphonic music pieces with rich annotations. However, the sizes of these three datasets are relatively small for training most deep generative models.
\end{spacing}

\subsection{Requirements of Datasets for Piano Arrangement}
We list the requirements of a music dataset suitable for the study of piano arrangement. The design objective of POP909 is to create a reliable, rich dataset that satisfies the following requirements.
\begin{itemize}[leftmargin = 10pt]
    \item \textbf{A style-consistent piano track}: The piano track can either be an re-orchestration of the original audio or an accompaniment of the lead melody. 
    \item \textbf{Lead melody or audio}: the necessary information for the arrangement task \circled{1} and \circled{2}, respectively.
    \item \textbf{Sufficient annotations} including key, beat, and chord labels. The annotations not only provide structured information for more controllable music generation, but also offer a flexible conversion between score and expressive performance.
    \item \textbf{Time alignment} among the piano accompaniment tracks, the lead melody or audio, and the annotations. 
    \item \textbf{A considerable size}: while traditional machine learning models can be trained on a relatively small dataset, deep learning models usually require a larger sample size (expected 50 hours in total duration).
\end{itemize}

% The design objective of POP909 is to create a reliable, rich dataset that satisfies the above requirements. It uniquely provides a stable piano arrangements in popular songs, which meets both instrument and genre requirement. The manually labeled lead melodies it has are unavailable in many datasets. Furthermore, the extracted annotations can make POP909 applicable to more tasks.

\section{Dataset Description}
\label{sec:3:datades}
POP909 consists of piano arrangements of 909 popular songs. The arrangements are time-aligned to the corresponding audios and maintain the original style and texture. Extra annotation includes beat, chord, and key information.

\subsection{Data Collection Process}\label{sec:3:1:collection}
We hire professional musicians to create piano arrangements. In order to maintain a high-quality standard of the arrangements, we divide the musicians into two teams: the \textit{arranger team} and the \textit{reviewer team}. The collection is finalized through an iterative procedure between two teams. For each song, each iteration goes through three steps: 
\begin{enumerate}[leftmargin = 10pt]
    \item Arrangement: the arranger team creates an arrangement from scratch, or revise the previous version of arrangement.
    \item Review: the reviewer team decides whether the current version is qualified and comments on how to improve the arrangement in case further revisions are required.
    \item Discussion: musicians from both teams catch up with the progress, discuss and improve details of arrangement standards.
\end{enumerate}

We start the process from a list of 1000 popular songs and finally select 909 songs with high arrangement quality. We not only present the last revision (i.e., the qualified version) of each song but also provide the unqualified versions of each song created during the iterative process. This multi-version feature may potentially offer a broader application scenario of the dataset.

% need to calc here
\subsection{Data Content and Format}
In POP909, the total duration of 909 arrangements is about 60 hours. The songs are composed by 462 artists. The release of all songs spans around 60 years (from the earliest in 1950s to the latest around 2010). % Most songs contain 4 - 5 versions of arrangements.

\begin{figure}
    \centering
    \includegraphics[width = \columnwidth]{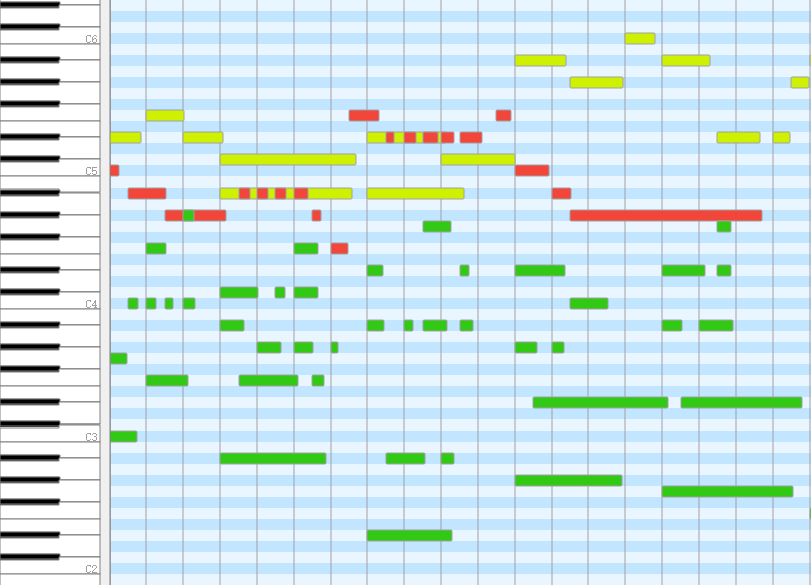}
    \caption{An example of the MIDI file in a piano roll view. Different colors denote different tracks (red for MELODY, yellow for BRIDGE, and green for PIANO).}
    \label{fig:midi_track}
\end{figure}

Each piano arrangement is stored in MIDI format with three tracks.
% The derived annotations are stored in text format, which contains the changes in beat, key and chord labels within the timeline.
Figure \ref{fig:midi_track} shows an example of a three-track MIDI file, in which different tracks are labeled with different colors. The three tracks are:
\begin{itemize}[leftmargin = 10pt]
    \item MELODY: the lead (vocal) melody transcription.
    \item BRIDGE: the arrangement of secondary melodies or lead instruments.
    \item PIANO: the arrangement of main body of the accompaniment, including broken chords, arpeggios, and many other textures.
\end{itemize}
Here, the combination of BRIDGE and PIANO track forms the piano accompaniment arrangement of the original song. Each MIDI file is aligned with the original audio by manually labeled tempo curve. Moreover, each note event contains expressive dynamics (i.e., detailed velocity control) based on the original audio. 

Beat, chord, and key annotations are provided in five separate text files for each song. Annotations for beat and chord have both MIDI and audio versions while key changes annotations are merely extracted from audios.\!\footnote{For annotations from MIDI files, the qualified (final) version of arrangements is used.} The relevant music information retrieval algorithms are discussed in Section~\ref{sec:4:annotation}.

%For each song, we provide three types of annotations in five text files: beat and downbeat positions extracted from both MIDI and audios, chord annotations extracted from both MIDI and audios, and key changes annotations extracted from audios.\!\footnote{The MIDI arrangements used in annotations extraction refer to the qualified (final) version.} All the annotations are extracted by various music information retrieval algorithms discussed in Section~\ref{sec:4:annotation}. 

% 这段没啥用，method就直白说数据里有啥。数据集的香在related work里说很清楚了。
% Overall, POP909 is a comprehensive dataset for music arrangement in Task 2 since we follow the timing of original audios to arrange the music. In addition, with the help of MIR algorithms and some manual labels, POP909 can be further used in Task 1. The annotations are not perfect, still, they are very useful in various generation tasks.

% Although the annotations may not be accurate enough for the purpose of professional music , we show in the experiments that they can still be used for generation tasks.)
% \kechen{[I suggest to rephrase as "Although annotations cannot achieve the retrieval level accuracy, we show in Section 5 that they can be used for generation tasks.]} % 不是cannot achieve，是可能didn't achieve。。。 但我们不知道能不能。

\begin{figure*}[t]
    \centering
    \includegraphics[width = 12cm]{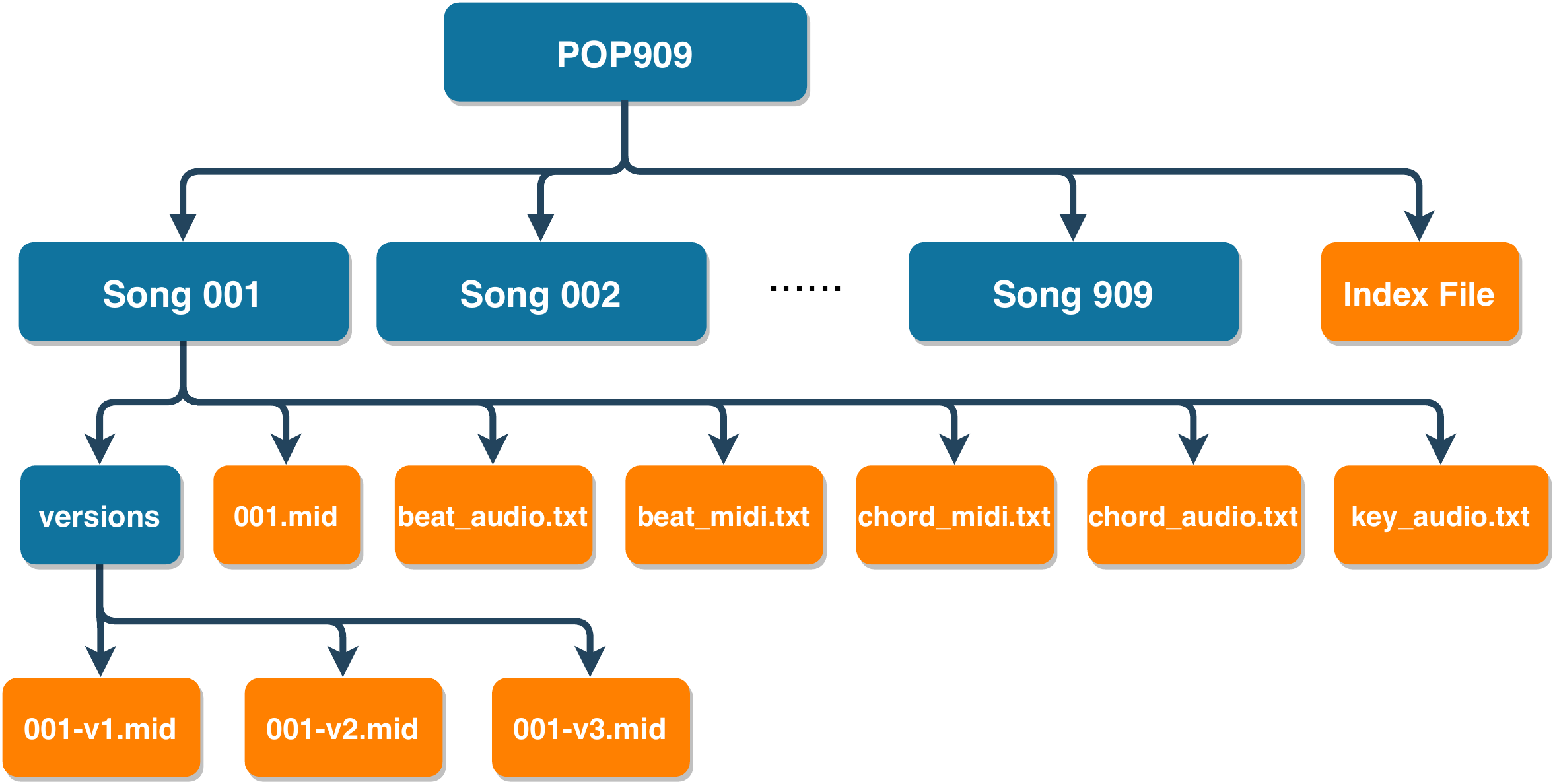}
    \caption{The folder structure of POP909. The blue boxes denote the folder and the orange boxes denote the file.}
    \label{fig:pop_structure}
\end{figure*}

\subsection{Data Folder Structure}
% In this subsection, we will go over the folder structure of POP909 in the released version.
Figure \ref{fig:pop_structure} demonstrates the folder structure of POP909. In the root directory, there are 909 folders, corresponding to 909 songs. In each folder, we provide the MIDI format arrangement, text format annotations, and a folder of all arrangement versions produced during the iterative processes.
% gus comments ends here 

The annotation files contain beat, chord and key annotations in plain text format. Table \ref{tab:annotation-example} shows the partial annotations of the song \texttt{003} in table format for better illustration purposes. For the beat annotation, \texttt{beat\_audio} and \texttt{beat\_midi} are the annotation files extracted from audio and MIDI, respectively. The source of chord and key annotations are indicated in a similar way. 

Finally, we provide an index file in the root directory containing the song name, artist name, number of modified times and other useful metadata of the dataset.

\section{Annotation Methods}
\label{sec:4:annotation}
In this section, we discuss how we annotate the beat, chord and key information. For each of the three tasks, different algorithms are applied to extract information from MIDI or audio. 

\subsection{Beat \& Downbeat Estimation}
We first extract beat information from MIDI files by taking advantage of two features of the MIDI performance: (1) human-annotated tempo curves, and (2) the accompaniment figure of arrangements which shows a significant sign of beat and downbeat attacks. 

Our method can be seen as a modification of the beat-tracking algorithms used in \cite{raffel2014intuitive, grohganz2014estimating}. First, we estimate the initial beat position and use the tempo curve to deduce subsequent beat positions. Second, we estimate the number of beats in a measure by calculating the auto-correlation of the extracted beat features (MIDI onset and velocity), assuming time signature is in general consistent within one song except for some infrequent phase changes. Finally, we search among all the possible phase shifts and find the optimal beat track that has the highest correlation with the extracted features. % Finally, we estimate in each beat, whether a beat consists of duplets or triplets. 

We also provide the beat and downbeat annotations extracted from the audio using the algorithm introduced in \cite{bock2016joint} and compare them with the annotations extracted from MIDI.  

For beat position estimation, the two algorithms have more than 90\% consistency when the maximum error tolerance is 100 ms, which is acceptable in the data collection process. For downbeat estimation, the two algorithms have 80\% agreement. We provide both extraction results in our annotation files.

\subsection{Chord Label Extraction}
We also provide the chord labels extracted from both MIDI and audio files. For the audio chord recognition, we adopt a large-vocabulary chord transcription algorithm by \cite{jiang2019largevocabulary}. As chord changes in popular music are most likely to happen at beat positions, we post-process the chord boundaries by aligning them to beats to produce the final chord labels.

\begin{table}[H]
\centering
{\small
\begin{tabular}{@{}cccc@{}}
\toprule
% \multicolumn{4}{c}{\textbf{Annotation Files of Song \texttt{003}}} \\ 
% \midrule\midrule
 \small{\textbf{file}} &\small{\textbf{beat time}} & \small{\textbf{downbeat\_1}} & \small{\textbf{downbeat\_2}} \\ \midrule
\texttt{beat\_midi} & \begin{tabular}[c]{@{}c@{}}0.02\\ 0.75\\ 1.49 \\ 2.22\\2.95\\3.68 \\... \end{tabular} & \begin{tabular}[c]{@{}c@{}}1.0\\ 0.0\\ 1.0\\0.0\\1.0\\0.0\\...\end{tabular} & \begin{tabular}[c]{@{}c@{}}0.0\\ 0.0\\ 1.0\\0.0\\0.0\\0.0\\...\end{tabular} \\ \midrule\midrule
 \small{\textbf{file}} & \small{\textbf{beat time}} & \multicolumn{2}{c}{\small{\textbf{beat order}}} \\ \midrule
\texttt{beat\_audio} & \begin{tabular}[c]{@{}c@{}}1.46\\ 2.18\\ 2.92\\3.66\\...\end{tabular} & \multicolumn{2}{c}{\begin{tabular}[c]{@{}c@{}}1.0\\ 2.0\\ 3.0\\4.0\\...\end{tabular}} \\ \midrule\midrule
\small{\textbf{file}} & \small{\textbf{start time}} & \small{\textbf{end time}} & \small{\textbf{chord}} \\ \midrule
\texttt{chord\_midi} & \begin{tabular}[c]{@{}c@{}}0.02\\ 0.75\\ 1.49\\ 4.41\\...\end{tabular} & \begin{tabular}[c]{@{}c@{}}0.75\\1.49\\4.41\\7.34\\...\end{tabular} & \begin{tabular}[c]{@{}c@{}}\texttt{N}\\ \texttt{N}\\ \texttt{G:min7}\\ \texttt{Eb:sus2} \\...\end{tabular} \\ \midrule\midrule
\small{\textbf{file}} & \small{\textbf{start time}} & \small{\textbf{end time}} & \small{\textbf{chord}} \\ \midrule
\texttt{chord\_audio} & \begin{tabular}[c]{@{}c@{}}0.00\\ 2.46\\ 4.39\\ ...\end{tabular} & \begin{tabular}[c]{@{}c@{}}1.46\\ 4.39\\ 7.31\\...\end{tabular} & \begin{tabular}[c]{@{}c@{}}\texttt{N}\\ \texttt{G:min7}\\ \texttt{Eb:maj(9)}\\...\end{tabular} \\ \midrule\midrule
\small{\textbf{file}} & \small{\textbf{start time}} & \small{\textbf{end time}} & \small{\textbf{key}} \\ \midrule
\texttt{key\_audio} & \begin{tabular}[c]{@{}c@{}}1.46\end{tabular} & \begin{tabular}[c]{@{}c@{}}226.00\end{tabular} & \begin{tabular}[c]{@{}c@{}}\texttt{Bb:maj}\end{tabular} \\ %\midrule\midrule
\toprule

\end{tabular}
}
\caption{The first several lines of the annotation files for song \texttt{003}. ``downbeat\_1'' and ``downbeat\_2'' in \texttt{beat\_midi} are the two downbeat extractors under simple meter and compound meter assumptions, respectively.}
\label{tab:annotation-example}
\end{table}

For MIDI chord recognition, we adopt a method similar to the one proposed in \cite{pardo2002algorithms}. We made two minor changes based on the original algorithm. First, the chord segmentation is performed on the beat level. Second, we alter the chord templates to include more chord qualities used by pop songs: (1) triads (\texttt{maj}, \texttt{min}, \texttt{dim}, \texttt{aug}) with inversions, (2) basic sevenths (\texttt{maj7}, \texttt{min7}, \texttt{7}, \texttt{dim7}, \texttt{hdim7}) with inversions, (3) suspended chords (\texttt{sus2}, \texttt{sus4}, \texttt{sus4(b7)}), and (4) sixth chords (\texttt{maj6}, \texttt{min6}).

Note that the arrangement and its original audio may have different chord progressions. For example, a \texttt{C:maj} chord may be arranged into \texttt{C:sus2}, if necessary. Therefore, both annotations are reasonable and they are not necessarily consistent with each other. To compare the extraction accuracy, we compute the matching rate of the root notes of the chords extracted from both methods. Results show that the matching degree of more than 800 songs in POP909 are above 75\%. On the other hand, there are still a few songs whose matching degrees are below 40\%. The main reasons are: (1) some of these audio recordings are slightly out of tune, and (2) some parts of the audio have complicated sound effects, in which case our teams decide to re-arrange the chord progression. 

\subsection{Key Signature Extraction}
We also provide key signature annotation from the audio files. We adopt an algorithm very similar to \cite{korzeniowski2017end}. The original algorithm performs the key classification for a whole song based on the averaged frame-wise feature. In our modified algorithm, we also allow key changes in the middle of the song using a median filter to post-process the frame-level labels.

\section{Experiments}
\label{sec:5:expr}
In this section, we conduct two baseline  experiments on music (score-modality) generation with the POP909 dataset: 1) polyphonic music generation (without melody condition), % flow based on the previous music context (e.g., 4-measure past context); 
and 2) piano arrangement generation conditioned on melody. For both tasks, we use the Transformer architecture \cite{tfm} for its advantages in capturing long-term dependencies on time-series data. %Specifically, our model is designed based on the GPT-2 framework \cite{gpt2} with relative positional encoding used in 

% \begin{figure}[t]
%     \centering
%     \includegraphics[width = \columnwidth]{figs/figure_6.PNG}
%     \caption{Matching degree of chord extraction in POP909. More than 800 songs in POP909 are above 75\%, but there are still a few songs below 40\%. }
%     \label{fig:chord_accuracy}
% \end{figure}

\subsection{Polyphonic Music Generation}\label{sec:5:1}

% We decide to use Transformer \cite{tfm} due to its advantages in capturing long-term dependencies of time-series data. We build the model based on the GPT-2 framework \cite{gpt2} with relative positional encoding used in \cite{naacl-relative-att, musictransformer}. 

% \begin{figure*}	
% 	\centering
% 	\begin{subfigure}[t]{\columnwidth}
% 		\centering
% 		\includegraphics[width=\columnwidth]{figs/7a_f.png}
% 		\caption{Arabic numerals}\label{fig:1a}		
% 	\end{subfigure}
% 	\quad
% 	\begin{subfigure}[t]{\columnwidth}
% 		\centering
% 		\includegraphics[width=\columnwidth]{figs/7b_f.png}
% 		\caption{Arabic numerals}\label{fig:1b}
% 	\end{subfigure}
% 	\caption{Capital Roman numerals}\label{fig:1}
% \end{figure*}

\begin{figure*}[t]	
	\centering
	
	\begin{subfigure}[t]{17cm}
		\centering
		\includegraphics[width=\textwidth]{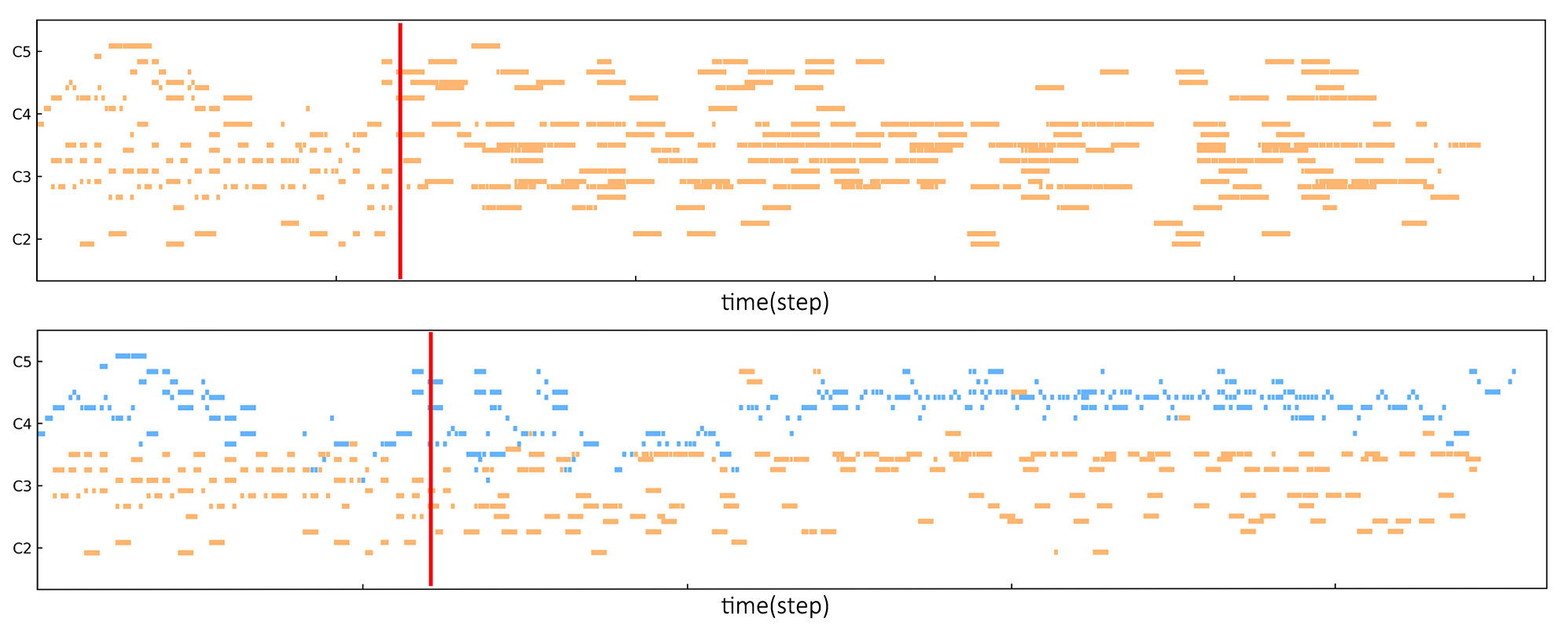}
		\caption{Generation results of example a.}\label{fig:mt_generation_a}		
	\end{subfigure}
	\quad
	\begin{subfigure}[h]{17cm}
		\centering
		\includegraphics[width=\textwidth]{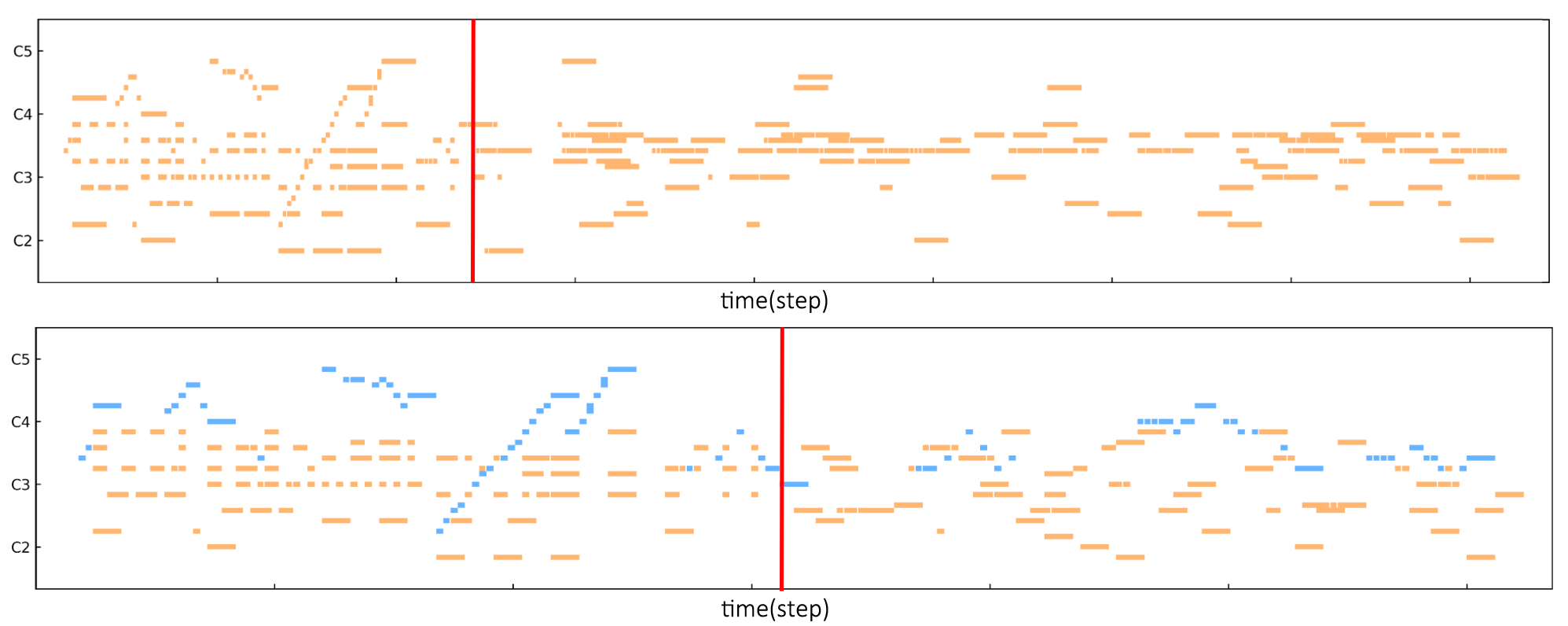}
		\caption{Generation results of example b.}\label{fig:mt_generation_b}
	\end{subfigure}
 	\caption{Generation examples with POP909 dataset. Unconditioned polyphonic music generation and piano arrangement generation (blue for the melody, orange for the accompaniment) of the two selected examples are displayed. }\label{fig:mt_generation}
	\quad
% 	\begin{subfigure}[b]{17cm}
% 		\centering
% 		\includegraphics[width=\textwidth]{figs/7c_1.png}
% 		\caption{Another two examples for the context-conditioned music generation and the music arrangement generation.}\label{fig:mt_generation_c}
% 	\end{subfigure}
% 	\caption{The example of generation via music arrangement model. Horizontal axis represents standard midi time steps.}\label{fig:mt_generation}
\end{figure*}

% \begin{figure*}[t]	
% 	\centering
	
% 	\begin{subfigure}[t]{17cm}
% 		\centering
% 		\includegraphics[width=\textwidth]{figs/7a.png}
% 		\caption{The complete generation given the previous context.}\label{fig:mt_generation_a}		
% 	\end{subfigure}
% 	\quad
% 	\begin{subfigure}[h]{17cm}
% 		\centering
% 		\includegraphics[width=\textwidth]{figs/7b.png}
% 		\caption{The arrangement generation (blue for the melody, orange for the accompaniment).}\label{fig:mt_generation_b}
% 	\end{subfigure}
% 	\quad
% 	\begin{subfigure}[b]{17cm}
% 		\centering
% 		\includegraphics[width=\textwidth]{figs/7c_1.png}
% 		\caption{Another two examples for the context-conditioned music generation and the music arrangement generation.}\label{fig:mt_generation_c}
% 	\end{subfigure}
% 	\caption{The example of generation via music arrangement model. Horizontal axis represents standard midi time steps.}\label{fig:mt_generation}
% \end{figure*}

% \subsubsection{Data Processing}
We use a transformer encoder with relative positional encoding \cite{naacl-relative-att, musictransformer} to model the distribution of polyphonic music. We adopt a MIDI-like event-based representation slightly modified from \cite{performance-rnn,musictransformer} to encode the polyphonic music. Each piece of music is represented as a series of events, including note onsets, offsets, velocity changes, and time shifts. We further quantize time shifts tokens under the resolution of $1 \over 4$ beat. In total, we use 16 time-shift events, ranging from $1 \over 4$ beat to 4 beats. Longer notes or rests can be represented by multiple time-shift tokens in a sequence. Table \ref{tab:tokenization} shows the details of our data representation. 

\begin{table}[]
\center
\begin{tabular}{|c|c|}
\hline
\textbf{Event type} & \textbf{Tokenization} \\ \hline
Note-On & \begin{tabular}[c]{@{}c@{}}0-127 (MELODY \& BRIDGE track)\\ 256-383 (PIANO track)\end{tabular} \\ \hline
Note-Off & \begin{tabular}[c]{@{}c@{}}128-255 (MELODY \& BRIDGE)\\ 384-511 (PIANO track)\end{tabular} \\ \hline
Time-Shift & 512-527 \\ \hline
Velocity & 528-560 \\ \hline
\end{tabular}
\caption{The tokenization of the modified MIDI-like event sequence representation.}
\label{tab:tokenization}
\end{table}

\begin{table}[]
\center
\begin{tabular}{@{}cccc@{}}
\toprule
\multicolumn{4}{c}{GPT-2-based transformer in POP909} \\ \midrule
Train Loss & Train Acc. & Test Loss & Test Acc. \\
2.08978 & 0.62021 & 2.38122 & 0.54529 \\ \bottomrule
\end{tabular}
\caption{The report of  training and test loss and prediction accuracy of MIDI event tokens.}
\label{tab:mt_loss}
\end{table}
    
% \subsubsection{Training Scheme}
% We feed the MIDI-event like data representation into a GPT-2 transformer structure with a mask blocking the future context. In addition, we embed both absolute and relative positional encoding into the input. The training procedure follows the next word prediction task (i.e., use $x_1$ to predict $x_2$, $\{x_1,x_2\}$ to predict $x_3$, $\{x_1,x_2,x_3\}$ to predict $x_4$, so on so forth). The model acts as an auto-regressive model in the generation phase. 

We split the dataset into 3 subsets: 90\% for training, 5\% for validation, and 5\% for testing. We set the maximum sequence length $L = 2048$, transformer hidden size $H = 512$, the number of attention heads $h = 6$, and the number of attention layers $N = 6$. Cross Entropy loss is used as the loss function and early stopping is applied.
%to find better but less over-fitting results. 

We use Adam optimizer \cite{adam} with hyperparameters $\beta_1 = 0.9,\beta_2 = 0.998$. We further adopt the warm-up schedule to  control the learning rate. Formally, at the $i$-th warm-up step, the learning rate%the hidden dimension size $d$, and the hyperparameter $S = 4000$: 
\begin{equation}
    lr = \frac{1}{\sqrt H} \ \times \min(\frac{1}{\sqrt i}\ , \frac{i}{S \sqrt S} ) \text{,}
\end{equation}
where $S = 4000$ is a hyperparameter controlling the number of warm-up steps. The training  result is presented in Table~\ref{tab:mt_loss}. 
% \kechen{Sorry it's the table, not the figure}

\subsection{Piano Arrangement Generation}
In the second experiment, we design an automatic piano arrangement task: piano accompaniment generation conditioned on the melody. In the data processing step, we first merge the MELODY track and the BRIDGE track into the \textit{main melody} and regard PIANO track the \textit{piano accompaniment}.

We use the same (trained) model in Section~\ref{sec:5:1} to model the joint distribution of the main melody and piano accompaniment. During the inference, we force the generated melody to match the given melody condition, generating the most likely accompaniment conditioned on the melody. (A similar conditional generation method has been used in \cite{lakhnes}.)

\subsection{Experiment Results}
Figure \ref{fig:mt_generation} shows several examples generated by the trained model. In each subfigure, the top piano roll shows the polyphonic music generation (introduced in Section 5.1) result and the bottom piano roll shows the piano arrangement generation (introduced in Section 5.2) result conditioned on the main melody (the blue track). In both cases, the first 500 MIDI-event tokens are given as the context; the red line separates the given context and the generated outputs. 
%我改到这儿，不知far from practice 是什么意思
We see that the generated pieces capture basic harmonic relationships between the melody and accompaniment and contain consistent rhythmic patterns. Although the quality is still far from the music generated by state-of-the-art algorithms \cite{musictransformer, polydis}, they serve as a baseline to illustrate our dataset usage. We believe that the model can produce better and more structured results with the development of deep generative models.

\section{Conclusion}
\label{sec:6:conclusion}
In conclusion, we contributed POP909, a tailored dataset for music arrangement. It contains multiple versions of professional piano arrangements in MIDI format of 909 popular songs, together with precise tempo curve aligned to the original audio recordings. We also provide annotations of tempo, beat, downbeat, key, and chord labels. To guarantee a high data quality, the dataset was collected via the collaboration of two groups of professional musicians, arrangers and reviewers, in an interactive process. Apart from the arrangement problem, the POP909 dataset serves as a high-quality resource for structural music generation and cross-modal music generation.

% \section{Acknowledgement}
% There is nothing we can write about the acknowledgement.
% We expect that POP909 can make relevant contributions to the task of automatic music arrangement from the perspective of the dataset.

% The motivation behind it is to support for the music arrangement generation model, so as to realize the model's greater parameterized potential. The unity of genre and instrumentation is the most basic requirement of this dataset. We use manual methods to collect and refine the data, and implement latest MIR algorithms to increase the universality of the dataset. POP909, therefore, can not only be a professional musical performance data, its annotations also provide the support of musical score data. Both can play a huge role in the music generation model. We hope that POP909 will attract more attentions from researchers to work on music arrangement generation, a specific but rising topic. To this end, we conduct automatic music arrangement baselines to show its availability and flexibility in generative tasks.   

\bibliography{ISMIRtemplate}

\begin{thebibliography}{10}
\providecommand{\url}[1]{#1}
\csname url@samestyle\endcsname
\providecommand{\newblock}{\relax}
\providecommand{\bibinfo}[2]{#2}
\providecommand{\BIBentrySTDinterwordspacing}{\spaceskip=0pt\relax}
\providecommand{\BIBentryALTinterwordstretchfactor}{4}
\providecommand{\BIBentryALTinterwordspacing}{\spaceskip=\fontdimen2\font plus
\BIBentryALTinterwordstretchfactor\fontdimen3\font minus
  \fontdimen4\font\relax}
\providecommand{\BIBforeignlanguage}[2]{{%
\expandafter\ifx\csname l@#1\endcsname\relax
\typeout{** WARNING: IEEEtran.bst: No hyphenation pattern has been}%
\typeout{** loaded for the language `#1'. Using the pattern for}%
\typeout{** the default language instead.}%
\else
\language=\csname l@#1\endcsname
\fi
#2}}
\providecommand{\BIBdecl}{\relax}
\BIBdecl

\bibitem{mysong}
I.~Simon, D.~Morris, and S.~Basu, ``Mysong: automatic accompaniment generation
  for vocal melodies,'' in \emph{Proceedings of the 2008 Conference on Human
  Factors in Computing Systems ({CHI})}.\hskip 1em plus 0.5em minus 0.4em\relax
  Florence, Italy: {ACM}, 2008, pp. 725--734.

\bibitem{elowsson}
A.~Elowsson and A.~Friberg, ``Algorithmic composition of popular music,'' in
  \emph{the 12th International Conference on Music Perception and Cognition and
  the 8th Triennial Conference of the European Society for the Cognitive
  Sciences of Music}, 2012, pp. 276--285.

\bibitem{wang2018framework}
Z.~Wang and G.~Xia, ``A framework for automated pop-song melody generation with
  piano accompaniment arrangement,'' \emph{arXiv preprint arXiv:1812.10906},
  2018.

\bibitem{musegan}
H.~Dong, W.~Hsiao, L.~Yang, and Y.~Yang, ``Musegan: Multi-track sequential
  generative adversarial networks for symbolic music generation and
  accompaniment,'' in \emph{Proceedings of the Thirty-Second {AAAI} Conference
  on Artificial Intelligence}.\hskip 1em plus 0.5em minus 0.4em\relax {AAAI}
  Press, 2018, pp. 34--41.

\bibitem{audioreduction}
H.~Takamori, T.~Nakatsuka, S.~Fukayama, M.~Goto, and S.~Morishima,
  ``Audio-based automatic generation of a piano reduction score by considering
  the musical structure,'' in \emph{International Conference on Multimedia
  Modeling ({ICMM})}.\hskip 1em plus 0.5em minus 0.4em\relax Springer, 2019,
  pp. 169--181.

\bibitem{audioreduction2}
G.~Percival, S.~Fukayama, and M.~Goto, ``Song2quartet: A system for generating
  string quartet cover songs from polyphonic audio of popular music.'' in
  \emph{Proceedings of the 16th International Society for Music Information
  Retrieval Conference ({ISMIR})}, 2015, pp. 114--120.

\bibitem{hung2019musical}
Y.-N. Hung, I.~Chiang, Y.-A. Chen, Y.-H. Yang \emph{et~al.}, ``Musical
  composition style transfer via disentangled timbre representations,''
  \emph{arXiv preprint arXiv:1905.13567}, 2019.

\bibitem{reduction1}
E.~Nakamura and S.~Sagayama, ``Automatic piano reduction from ensemble scores
  based on merged-output hidden markov model,'' in \emph{Proceedings of the
  41st International Computer Music Conference ({ICMC})}, 2015.

\bibitem{reduction2}
J.-L. Huang, S.-C. Chiu, and M.-K. Shan, ``Towards an automatic music
  arrangement framework using score reduction,'' \emph{ACM Transactions on
  Multimedia Computing, Communications, and Applications ({TOMM})}, vol.~8,
  no.~1, pp. 1--23, 2012.

\bibitem{mr}
M.~Xu, Z.~Wang, and G.~Xia, ``Transferring piano performance control across
  environments,'' in \emph{{IEEE} International Conference on Acoustics, Speech
  and Signal Processing ({ICASSP})}.\hskip 1em plus 0.5em minus 0.4em\relax
  Brighton, United Kingdom: {IEEE}, 2019, pp. 221--225.

\bibitem{xiaexpressive}
G.~Xia, ``Expressive collaborative music performance via machine learning,''
  Ph.D. dissertation, Carnegie Mellon University.

\bibitem{jeong2019graph}
D.~Jeong, T.~Kwon, Y.~Kim, and J.~Nam, ``Graph neural network for music score
  data and modeling expressive piano performance,'' in \emph{Proceedings of the
  36th International Conference on Machine Learning ({ICML})}, 2019, pp.
  3060--3070.

\bibitem{lakhmidi}
C.~Raffel, ``Learning-based methods for comparing sequences, with applications
  to audio-to-midi alignment and matching,'' \emph{PhD thesis, Columbia
  University}, 2016.

\bibitem{JSB-chorale}
N.~Boulanger{-}Lewandowski, Y.~Bengio, and P.~Vincent, ``Modeling temporal
  dependencies in high-dimensional sequences: Application to polyphonic music
  generation and transcription,'' in \emph{Proceedings of the 29th
  International Conference on Machine Learning ({ICML})}.\hskip 1em plus 0.5em
  minus 0.4em\relax icml.cc / Omnipress, 2012.

\bibitem{maestro}
C.~Hawthorne, A.~Stasyuk, A.~Roberts, I.~Simon, C.-Z.~A. Huang, S.~Dieleman,
  E.~Elsen, J.~Engel, and D.~Eck, ``Enabling factorized piano music modeling
  and generation with the {MAESTRO} dataset,'' in \emph{7th International
  Conference on Learning Representations ({ICLR})}, New Orleans, LA, USA, 2019.

\bibitem{crestmuse}
M.~Hashida, T.~Matsui, and H.~Katayose, ``A new music database describing
  deviation information of performance expressions,'' in \emph{Proceedings of
  9th International Conference on Music Information Retrieval ({ISMIR})},
  Philadelphia, PA, USA, 2008, pp. 489--494.

\bibitem{rwcpop}
M.~Goto, H.~Hashiguchi, T.~Nishimura, and R.~Oka, ``{RWC} music database:
  Popular, classical and jazz music databases,'' in \emph{Proceedings of 3rd
  International Conference on Music Information Retrieval ({ISMIR})}, Paris,
  France, 2002.

\bibitem{nottingham}
E.~Foxley, ``Nottingham database,''
  \url{https://ifdo.ca/~seymour/nottingham/nottingham.html}, 2011.

\bibitem{musictransformer}
C.~A.~H. et~al., ``Music transformer: Generating music with long-term
  structure,'' in \emph{7th International Conference on Learning
  Representations ({ICLR})}, New Orleans, LA, USA, 2019.

\bibitem{lakhnes}
C.~Donahue, H.~H. Mao, Y.~E. Li, G.~W. Cottrell, and J.~J. McAuley, ``Lakhnes:
  Improving multi-instrumental music generation with cross-domain
  pre-training,'' in \emph{Proceedings of the 20th International Society for
  Music Information Retrieval Conference ({ISMIR})}, Delft, The Netherlands,
  2019, pp. 685--692.

\bibitem{chen2019effect}
K.~Chen, W.~Zhang, S.~Dubnov, G.~Xia, and W.~Li, ``The effect of explicit
  structure encoding of deep neural networks for symbolic music generation,''
  in \emph{2019 International Workshop on Multilayer Music Representation and
  Processing ({MMRP})}.\hskip 1em plus 0.5em minus 0.4em\relax IEEE, 2019, pp.
  77--84.

\bibitem{dai2018music}
S.~Dai, Z.~Zhang, and G.~G. Xia, ``Music style transfer: A position paper,''
  \emph{Proceeding of International Workshop on Musical Metacreation ({MUME})},
  2018.

\bibitem{deepbach}
G.~Hadjeres, F.~Pachet, and F.~Nielsen, ``Deepbach: a steerable model for bach
  chorales generation,'' in \emph{Proceedings of the 34th International
  Conference on Machine Learning, ({ICML})}.\hskip 1em plus 0.5em minus
  0.4em\relax Sydney, NSW, Australia: {PMLR}, 2017, pp. 1362--1371.

\bibitem{musicinpaint}
A.~Pati, A.~Lerch, and G.~Hadjeres, ``Learning to traverse latent spaces for
  musical score inpainting,'' in \emph{Proceedings of the 20th International
  Society for Music Information Retrieval Conference ({ISMIR})}, Delft, The
  Netherlands, 2019, pp. 343--351.

\bibitem{folkrnn}
B.~L. Sturm, J.~F. Santos, O.~Ben{-}Tal, and I.~Korshunova, ``Music
  transcription modelling and composition using deep learning,'' in
  \emph{Proceedings of 1st Conference on Computer Simulation of Musical
  Creativity ({CSMC})}, 2016.

\bibitem{ICSC-Ke}
K.~Chen, G.~Xia, and S.~Dubnov, ``Continuous melody generation via disentangled
  short-term representations and structural conditions,'' in \emph{14th
  International Conference on Semantic Computing ({ICSC})}.\hskip 1em plus
  0.5em minus 0.4em\relax San Diego, CA, USA: {IEEE}, 2020, pp. 128--135.

\bibitem{wavenet}
A.~van den Oord~et al., ``Wavenet: {A} generative model for raw audio,'' in
  \emph{The 9t Speech Synthesis Workshop}.\hskip 1em plus 0.5em minus
  0.4em\relax Sunnyvale, CA, USA: {ISCA}, 2016, p. 125.

\bibitem{jukebox}
P.~D. et~al., ``Jukebox: {A} generative model for music,'' \emph{CoRR}, vol.
  abs/2005.00341, 2020.

\bibitem{e-piano}
``International piano-e-competition,''
  \url{http://www.piano-e-competition.com/}.

\bibitem{raffel2014intuitive}
C.~Raffel and D.~P. Ellis, ``Intuitive analysis, creation and manipulation of
  midi data with pretty midi,'' in \emph{Proceedings of the 15th International
  Society for Music Information Retrieval Conference ({ISMIR}), Late Breaking
  and Demo Papers}, Taipei, Taiwan, 2014, pp. 84--93.

\bibitem{grohganz2014estimating}
H.~Grohganz, M.~Clausen, and M.~M{\"{u}}ller, ``Estimating musical time
  information from performed {MIDI} files,'' in \emph{Proceedings of the 15th
  International Society for Music Information Retrieval Conference ({ISMIR})},
  Taipei, Taiwan, 2014, pp. 35--40.

\bibitem{bock2016joint}
S.~B{\"{o}}ck, F.~Krebs, and G.~Widmer, ``Joint beat and downbeat tracking with
  recurrent neural networks,'' in \emph{Proceedings of the 17th International
  Society for Music Information Retrieval Conference ({ISMIR})}, New York City,
  United States, 2016, pp. 255--261.

\bibitem{jiang2019largevocabulary}
J.~Jiang, K.~Chen, W.~Li, and G.~Xia, ``Large-vocabulary chord transcription
  via chord structure decomposition,'' in \emph{Proceedings of the 20th
  International Society for Music Information Retrieval Conference ({ISMIR})},
  Delft, The Netherlands, 2019, pp. 644--651.

\bibitem{pardo2002algorithms}
B.~Pardo and W.~P. Birmingham, ``Algorithms for chordal analysis,''
  \emph{Computer Music Journal}, vol.~26, no.~2, pp. 27--49, 2002.

\bibitem{korzeniowski2017end}
F.~Korzeniowski and G.~Widmer, ``End-to-end musical key estimation using a
  convolutional neural network,'' in \emph{25th European Signal Processing
  Conference ({EUSIPCO})}.\hskip 1em plus 0.5em minus 0.4em\relax Kos, Greece:
  {IEEE}, 2017, pp. 966--970.

\bibitem{tfm}
A.~Vaswani, N.~Shazeer, N.~Parmar, J.~Uszkoreit, L.~Jones, A.~N. Gomez,
  L.~Kaiser, and I.~Polosukhin, ``Attention is all you need,'' in
  \emph{Advances in Neural Information Processing Systems 30: Annual Conference
  on Neural Information Processing Systems ({NeurIPS})}, 2017, pp. 5998--6008.

\bibitem{naacl-relative-att}
P.~Shaw, J.~Uszkoreit, and A.~Vaswani, ``Self-attention with relative position
  representations,'' in \emph{Proceedings of the 2018 Conference of the North
  American Chapter of the Association for Computational Linguistics: Human
  Language Technologies, NAACL-HLT ({NAACL})(Short Papers)}.\hskip 1em plus
  0.5em minus 0.4em\relax New Orleans, Louisiana, USA: Association for
  Computational Linguistics, 2018, pp. 464--468.

\bibitem{performance-rnn}
I.~Simon and S.~Oore, ``Performance rnn: Generating music with expressive
  timing and dynamics,'' \url{https://magenta.tensorflow.org/performance-rnn},
  2017.

\bibitem{adam}
D.~P. Kingma and J.~Ba, ``Adam: {A} method for stochastic optimization,'' in
  \emph{Proceeding of the 3rd International Conference on Learning
  Representations ({ICLR})}, San Diego, CA, USA, 2015.

\bibitem{polydis}
Z.~Wang, D.~Wang, Y.~Zhang, and G.~Xia, ``Learning interpretable representation
  for controllable polyphonic music generation,'' in \emph{Proceedings of 21st
  International Conference on Music Information Retrieval ({ISMIR}), virtual
  conference}, 2020.

\end{thebibliography}

% For non bibtex users:
%\begin{thebibliography}{citations}
%
%\bibitem {Author:00}
%E. Author.
%``The Title of the Conference Paper,''
%{\it Proceedings of the International Symposium
%on Music Information Retrieval}, pp.~000--111, 2000.
%
%\bibitem{Someone:10}
%A. Someone, B. Someone, and C. Someone.
%``The Title of the Journal Paper,''
%{\it Journal of New Music Research},
%Vol.~A, No.~B, pp.~111--222, 2010.
%
%\bibitem{Someone:04} X. Someone and Y. Someone. {\it Title of the Book},
%    Editorial Acme, Porto, 2012.
%
%\end{thebibliography}

\end{document}